# Informing Robot Wellbeing Coach Design through Longitudinal Analysis of Human–AI Dialogue[*]

Keya Shah
kts5726@nyu.edu
SMART Lab, New York University
Abu Dhabi, United Arab Emirates

Himanshi Lalwani
hl3937@nyu.edu
SMART Lab, New York University
Abu Dhabi, United Arab Emirates

Zein Mukhanov
zm2199@nyu.edu
SMART Lab, New York University
Abu Dhabi, United Arab Emirates

Hanan Salam
hs4461@nyu.edu
SMART Lab, New York University
Abu Dhabi, United Arab Emirates

## Abstract
Social robots and conversational agents are being explored as supports for wellbeing, goal-setting, and everyday self-regulation. While prior work highlights their potential to motivate and guide users, much of the evidence relies on self-reported outcomes or short, researcher-mediated encounters. As a result, we know little about the interaction dynamics that unfold when people use such systems in real-world contexts, and how these dynamics should shape future robot wellbeing coaches. This paper addresses this gap through content analysis of 4352 messages exchanged longitudinally between 38 university students and an LLM-based wellbeing coach. Our results provide a fine-grained view into how users naturally shape, steer, and sometimes struggle within supportive human-AI dialogue, revealing patterns of user-led direction, guidance-seeking, and emotional expression. We discuss how these dynamics can inform the design of robot wellbeing coaches that support user autonomy, provide appropriate scaffolding, and uphold ethical boundaries in sustained wellbeing interactions.

## ACM Reference Format:
Keya Shah, Himanshi Lalwani, Zein Mukhanov, and Hanan Salam. 2026. Informing Robot Wellbeing Coach Design through Longitudinal Analysis of Human–AI Dialogue. In *Companion Proceedings of the 21st ACM/IEEE International Conference on Human-Robot Interaction (HRI Companion '26), March 16–19, 2026, Edinburgh, Scotland, UK*. ACM, New York, NY, USA, 5 pages. https://doi.org/10.1145/3776734.3794409

[*]This work is supported by the NYUAD Center for Interdisciplinary Data Science & AI (CIDSAI), funded by Tamkeen under the NYUAD Research Institute Award CG016.

## 1 Introduction

The potential of robots to serve as coaches, companions, and vital supports in areas like wellbeing and self-regulation is rapidly being realized [20, 26, 27]. When a robot assumes a coaching role, its influence on users' autonomy, agency, emotional vulnerability, and decision-making becomes a central design concern [13, 18]. According to Self-Determination Theory (SDT) [36], sustained motivation, persistence and wellbeing are best achieved through autonomy-supportive interactions. Designing such autonomy-supportive robot coaches therefore requires a deep understanding not only of robot behavior, but also of how humans actually respond to guidance in naturalistic, multi-session support contexts.

Concurrently, Large Language Models (LLMs) are becoming the core technology for many future robot coaches [13]. In both standalone and robotic applications, LLMs support flexible natural-language coaching, including goal-setting and reflective dialogue [34, 38, 43]. While physical embodiment can shape user perception through social presence, authority, and affective cues [15, 18], it does not replace the underlying interaction logic that structures these conversations. Examining interaction dynamics in LLM-based wellbeing coaching therefore provides an important foundation for anticipating how these behaviors may be experienced when instantiated in embodied robot coaches. While prior work has documented how users disclose emotions to conversational agents [14, 25], we know far less about how people negotiate, resist, adapt, or comply with system-generated suggestions in multi-turn support contexts. These dynamics become ethically critical once the same behaviors are embodied in robots. This raises two questions:

**RQ1.** How do students' natural interactions with an LLM-based wellbeing coach reveal broader interaction dynamics in supportive human–AI dialogue?

**RQ2.** How can these interaction dynamics inform the social presence, ethical boundaries, and autonomy-supportive behaviors of future robot wellbeing coaches?

To investigate these, we analyzed 4,352 real-world messages exchanged between university students and an LLM-based wellbeing coach deployed over a week. Using quantitative content analysis, guided by established concepts of human autonomy, agency, and interaction found in psychological and Human-Robot Interaction (HRI) literature [6, 8, 13, 25, 33, 42], we coded how users responded to guidance, set or revised goals, pushed back, complied, and disclosed emotions. Our findings highlight the interaction dynamics that emerge even in text-only AI coaching, and highlight considerations essential for designing future robot coaches that are supportive without becoming directive or coercive.

## 2 Related Work

**AI and Social Robots for Wellbeing Support.** Socially assistive robots (SARs) are being used in wellbeing and mental health contexts to provide encouragement, structure, and supportive interaction without physical assistance. Studies show that such robots can guide positive psychology practices [2, 20], support behavior change interventions [19], and support psychosocial wellbeing by improving outcomes such as loneliness, social interaction, mood, and stress in nonclinical settings [31].

In parallel, conversational AI wellbeing tools provide accessible supports for emotional expression [7, 28], self-reflection [22], everyday stress management [41], and behavior therapy interventions [12]. University students frequently describe these systems as non-judgmental and easy to talk to, which can make it easier to articulate personal concerns or explore stressors [35, 39]. Systematic reviews similarly report improvements in perceived wellbeing and reductions in anxiety [29], and recent work links conversational AI use with higher emotional self-efficacy and perceived autonomy [40]. Together, these findings show that wellbeing support in both SARs and conversational AI is shaped through ongoing interaction with users. Understanding how these supportive interactions unfold is important for informing the design of future robot coaches.

**Interaction Dynamics.** Supportive interactions with AI, whether delivered through chatbots or SARs, are shaped by a collection of interaction dynamics that determine how people express themselves, respond to guidance, and share control in the moment [30]. Prior work has examined interaction dynamics such as human autonomy, agency, emotional disclosure, compliance, negotiation, and rumination [3, 5, 10, 13, 15, 17]. However, these studies typically treat each dynamic independently and within highly specific tasks. As a result, we know less about how these dynamics coexist and interact during open-ended conversations for wellbeing, where users shift fluidly between expressing emotion, revising goals, accepting or resisting suggestions, and returning to ongoing concerns.

Recent work in HRI highlights that protecting and promoting human autonomy and agency is an ethical and design-critical priority as both are prerequisites for intrinsic motivation and wellbeing [13]. Emotional self-disclosure, defined as the act of sharing one's emotional states, feelings, stressors [25, 33], also commonly occurs in interactions with artificial agents, as users often perceive conversational systems and robots as emotionally safe, nonjudgmental partners for expression [24, 25]. However, this might drift to rumination, where users repeatedly revisit the same worries or stressors [42]. Consequently, it is important to understand how wellbeing systems can encourage healthy disclosure while preventing it from becoming repetitive or maladaptive. Another important interaction dynamic is negotiation, understood as resolving differing preferences to reach mutually acceptable solutions [5, 6]. Negotiation and compliance are increasingly relevant as social robots appear in education, workplaces, and customer-service settings [5, 15, 17, 21]. Understanding how users negotiate and comply with AI systems is essential for ensuring that collaborative interactions remain balanced and do not unintentionally shift control toward the system.

## 3 Methodology

**Context.** The study examined natural use of an LLM-based wellbeing coach deployed to support university students with goal setting and everyday wellbeing. The coach was powered by the OpenAI GPT API and guided by a fixed system prompt instructing it to act as a supportive, non-clinical coach: asking open reflection questions, helping users refine and set goals, and suggesting practical strategies. To structure goal setting, the coach used the SMART framework [9], which encourages goals that are Specific, Measurable, Achievable, Relevant, and Time-bound, a format widely used to improve clarity and follow-through in wellbeing interventions. The prompt also incorporated light elements from the Acceptance and Commitment Therapy (ACT), a psychological intervention that promotes psychological flexibility [16]. The coach uses these elements only for values clarification and acceptance-oriented phrasing to support wellbeing and reflection. The system (see Figure 1) retained conversation history across sessions, enabling it to reference past goals and follow up on earlier concerns. It did not provide diagnostic, crisis, or therapeutic mental-health advice. Students accessed the system through a web interface and could initiate conversations at any time during the deployment period.

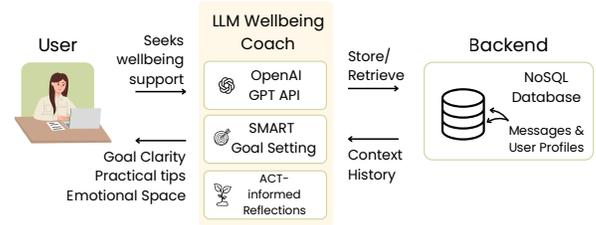

Figure 1: Overview of the LLM-based wellbeing coach system

**Participants.** 38 students participated in the one-week deployment. All participants were screened using the Patient Health Questionnaire (PHQ-9) [23]; only those scoring below 15 were invited to participate in the study to ensure ethical safety. The dataset contains 4,352 messages, with participants engaging in sustained multi-turn conversations averaging 116 total dialogue turns per participant, counting both student and system messages. Participation was voluntary. Students provided informed consent for their anonymized interaction logs to be used for research. Personally identifiable information was stripped before saving to database. This study was approved by the University Institutional Review Board.

**Protocol.** At onboarding, participants were told that the system was intended to help with everyday wellbeing and SMART goal setting. They were explicitly informed that the coach was not a therapeutic service and that it should not be used for clinical mental health needs. This framing was used to reduce over-reliance, clarify the limits of the system's capability, and discourage substitution of the tool for professional care. Researchers did not intervene during the study. Participants were free to stop using the tool at any time.

**Coding and Analysis.** We applied content analysis to characterize the interaction dynamics that emerged during use of the LLM wellbeing coach. Quantitative content analysis treats text as units that can be coded and counted to draw valid inferences [4]. The unit of analysis was each participant message, defined as one user turn in the dialogue. Assistant messages were reviewed for context but were not coded. Guided by frameworks from psychology and HRI [6, 8, 13, 25, 33, 42], we developed an interaction dynamics coding scheme that included autonomy, agency, negotiation, compliance, emotional disclosure and rumination (see Table 1). Messages could receive multiple codes when relevant. An additional coder reviewed the initial coding, and differences were resolved via discussion.

Table 1: Codebook for Content Analysis

| Interaction Dynamic | Definition | Examples From the User-AI Interactions |
|---|---|---|
| Autonomy | User expresses self-determined choices, preferences, or boundaries that guide the direction, pace, or content of the interaction [13]. | Coach: "Is there another area you'd like to focus on or chat about today?" P19: "Not really, I'm already feeling tired" |
| Agency | User initiates actions that drive the interaction, such as requesting resources, deciding next steps, or acting on prior guidance to shape the direction of the exchange [13]. | Coach: "Great progress today! Are you ready to wind down and enjoy the rest of your day?" P14: "No I want to keep working so that I don't feel stressed the rest of the week" |
| Emotional Self-Disclosure | User reveals feelings, vulnerabilities, affective states, or personal emotional context [24, 25, 33]. | Coach: "Hey! Good to see you. How are you feeling today?" P10: "I have no idea but I am so disspointed in myself so disappointed" |
| Rumination | User expresses repetitive, self-focused or problem-focused thinking that circles around the same concern or perceived failure [42]. | P22: "An interview of mine had gone bad and I'm really beating myself up about it" P22: "I'm still not able to get the interview out of my mind…this is interfering with my studying" |
| Negotiation | User modifies, adjusts, rejects, or collaboratively reshapes system suggestions or goals [5, 6]. | P27: "Hi! I think I set kind of unrealistic goals" Coach: "It's totally normal to adjust goals. Is there a part of your goal that feels unrealistic?" |
| Compliance | User explicitly accepts, agrees to, or commits to carrying out a suggestion proposed without substantial modification [8]. | Coach: "Opening the presentation alone is a win on heavy days. That still counts as progress." P20: "I started working, thanks for help" |

## 4 Findings

**Interaction Dynamics –** We report the distribution and frequency of the six interaction dynamics observed across participants.

*Autonomy.* Autonomy appeared in 37 of 38 participants (97.4%), with 1–14 instances per participant ($M = 6.1$ among those showing autonomy), and accounted for 10.5% of all user messages. It was typically expressed through users defining priorities or boundaries, for example, "I want to be the one who brings up results" and "I want a good balance between academics and social life," positioning themselves as primary decision-makers and the coach as a support rather than a directive authority.

*Agency.* Agency appeared in 37 of 38 participants (97.4%), with 1–46 instances per participant ($M = 12.1$ among those showing agency), and accounted for 20.8% of all user messages. It appeared when users took initiative to act independently or set next steps without prompting, for example, "I'll do this on my own," and "I'm going to start working now for the next couple of hours." Others explicitly managed task execution, such as, "I moved the goal check-up today to 9pm in my Google Calendar." These examples show users actively driving progress beyond the conversational space.

*Emotional Disclosure.* Emotional disclosure appeared in 26 of 38 participants (68.4%), with 1–12 instances per participant ($M = 2.8$ among those showing agency), and accounted for 3.3% of all user messages. Disclosure often involved concise but personal expressions of emotional state, such as "I've been feeling unseen lately," or "Feeling pretty anxious today." Some users also paired it with interpretation, for example, "I think it's the exhaustion from the day and the sadness." These utterances reflect moments where users shared internal emotional context to frame the interaction.

*Rumination.* Rumination appeared in 5 of 38 participants (13.2%), with 1–5 instances per participant ($M = 1.8$ among those who ruminated), and accounted for less than 1% of all user messages. When it did occur, it often involved self-focused reflection on perceived failure or distress. For example, one participant stated, "I feel like I could've performed better… I'm sure I've been rejected." followed later by "I'll basically be a failure." These sequences show rumination as looping concern rather than forward-oriented planning.

*Compliance.* Compliance appeared in 30 of 38 participants (78.9%), with 1–9 instances per participant ($M = 3.5$ among those who complied), and accounted for 4.9% of all user messages. It appeared as clear acceptance or endorsement of suggested actions, from brief confirmations such as "Yes, within 30 minutes of waking up sounds good" to reflective endorsements like "Hey, studied every topic today till now." These responses show moments where users aligned with the system's recommendations without modification.

*Negotiation.* Negotiation appeared in 30 of 38 participants (78.9%), with 1–13 instances per participant ($M = 4.2$ among those who negotiated), and accounted for 5.8% of all user messages. It occurred when users reshaped system suggestions to better fit their constraints or preferences, for example, "I aim for 180 mins of intense workout a week, not 30," or "I don't think I can finish my assignments by 9pm. It gives me pressure." These responses exemplify collaborative adjustment rather than disengagement.

*Overall pattern.* Agency and autonomy were the most pervasive interaction dynamics, together accounting for over 29% of all user utterances. Compliance and negotiation appeared in many users but at lower frequencies, and emotional disclosure appeared in more than half of participants. Rumination was rare and low in intensity. Overall, these patterns suggest that students often co-directed the wellbeing interaction, selectively accepted or adjusted guidance, and occasionally disclosed personal emotions, with minimal evidence of repetitive distress loops.

*Co-Occurrences in Interaction Dynamics.* Interaction dynamics rarely appeared in isolation; most participants exhibited multiple behaviors during the study. *Autonomy and agency* showed the strongest overlap, co-occurring in 36 of 38 participants (94.7%), indicating that users who set or adjusted goals also actively shaped conversation flow. *Negotiation* overlapped with both autonomy (76.3%) and agency (78.9%), suggesting that users who guided the interaction often refined, questioned, or modified system-generated suggestions. *Emotional disclosure* mostly co-occurred with autonomy and agency (65.8–68.4%), showing that reflective or vulnerable statements arose in conversations where users were already engaged and directive rather than passive. *Compliance* also co-occurred with autonomy and agency at high rates (76.3–78.9%), reinforcing that accepting guidance did not imply a lack of initiative; many participants both complied with suggestions and contributed direction or negotiation. Since compliance occurred infrequently overall, these co-occurrence patterns should be interpreted cautiously and understood as indicative rather than conclusive. *Rumination* exhibited minimal co-occurrence with any other construct (13.2%), indicating that repetitive distress-oriented exchanges were uncommon. This suggests that most wellbeing interactions were oriented toward action, adjustment, and goal-setting rather than repetitive distress.

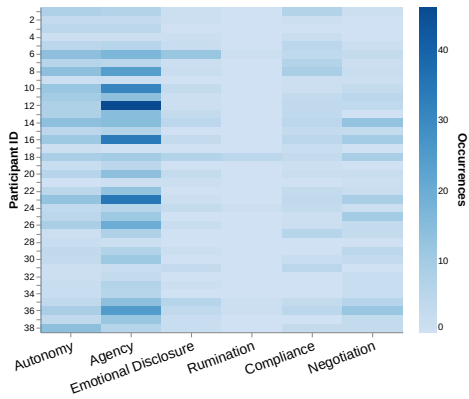

Figure 2: Interaction dynamics distribution per participant.

**Engagement-Related Behaviors –** Participants differed substantially in how often and how extensively they engaged with the wellbeing coach. Total user messages per participant ranged from 23 to 110 ($M = 56.71, SD = 22.16$), reflecting wide variation in interaction frequency. Message verbosity showed comparable dispersion, with average message length ranging from 1.45 to 20.95 words per message ($M = 6.42, SD = 4.48$). The moderate correlation between message frequency and verbosity ($r = .34$) indicates that these dimensions were only partially related; students who sent more messages did not necessarily write longer ones.

## 5 Discussion

From these interaction dynamics, we derive five design guidelines for robot wellbeing coaches.

***Prioritize user-led direction through shared initiative.*** The high prevalence of autonomy and agency indicates that students engage with supportive systems as collaborative partners rather than directive authorities, frequently initiating goals, steering topic transitions, and shaping dialogue pace. This pattern aligns with Self-Determination Theory, which identifies autonomy as central to intrinsic motivation and wellbeing [36], and with mixed-initiative interaction theory, which emphasizes flexible sharing of control during collaboration allowing goals to be refined through interaction rather than fully specified in advance [1]. Together, these findings motivate robot wellbeing coaches that support user-led direction through shared initiative while retaining the ability to intervene or redirect when safety or wellbeing concerns arise.

***Treat negotiation as engagement, not resistance.*** Negotiation emerged as a common interaction pattern as users refined, questioned, or adapted system-generated suggestions. Mixed-initiative interaction theory characterizes such behavior as part of collaborative coordination through comparing alternatives, refining proposals, and introducing constraints as interaction unfolds [1], a pattern reflected in our findings, where users negotiated by refining suggestions or articulating contextual constraints. These findings suggest that robot wellbeing coaches should treat negotiation as a signal of engagement [37], designing negotiation-aware interactions that adjust explanations, offer alternatives, or acknowledge constraints to avoid appearing rigid or prescriptive. Compliance must also be interpreted carefully. Although infrequent, its co-occurrence with autonomy and agency suggests that acceptance of guidance may, in some cases, coexist with active user participation rather than passive submission. Prior work shows that embodied robots can amplify perceived authority and social pressure to comply [15]. Robot wellbeing coaches should ensure that compliance reflects genuine choice rather than social pressure.

***Preserve user agency during emotional disclosure and calibrate social presence.*** Emotional disclosure mostly occurred when users were already exercising agency, suggesting that disclosure emerged when users felt in control of the interaction. Prior work shows that emotional disclosure is shaped by interactional roles and dominance, with user experience influenced more by the form of engagement than by conversational content alone [11]. In embodied systems, emotional behaviors and social presence cues in robots can increase mind attribution and perceived emotional competence, potentially leading users to infer therapeutic capacity beyond the system's actual capabilities [11, 15]. Together, these findings imply a dual design responsibility for robot wellbeing coaches: to preserve user control, which appears to enable functional disclosure, while avoiding signals of deeper emotional understanding that could foster over-reliance or overstep the system's role.

***Recognize and redirect early negative loops.*** Rumination was rare and less integrated with other interaction dynamics, though a small number of participants exhibited brief moments of repetitive negative focus. Clinical psychology defines rumination as a maladaptive pattern of sustained attention to distress that can intensify negative affect even when episodic rather than persistent [32, 42]. Thus, low frequency does not imply low design importance. In HRI, this concern is amplified by embodiment, as embodied agents can increase attentional focus and perceived legitimacy of system responses, potentially intensifying users' engagement with their emotional state [15]. Robot wellbeing coaches must therefore be able to detect and gently redirect emerging rumination through grounding or reframing strategies when it arises.

***Adapt to diverse engagement styles.*** Message length and frequency showed modest correlation, suggesting students use different engagement modes: rapid exchanges, extended reflective turns, or a mix of both. This suggests wellbeing support is not one-size-fits-all. Future robot coaches will need adaptive strategies that adjust pacing, prompting, and turn-taking to each user's style.

***Limitations and Conclusion.*** Our one-week deployment could not capture longer term effects such as habituation or evolving reliance. Because our participants were university students, the interaction patterns may not generalize to other groups. Our results suggest robot wellbeing coaches should respect user-led direction, adapt strategies to personal constraints, calibrate social presence, and support users without coercion.

## References

[1] James E Allen, Curry I Guinn, and Eric Horvtz. 1999. Mixed-initiative interaction. *IEEE Intelligent Systems and their Applications* 14, 5 (1999), 14–23.

[2] Minja Axelsson, Nikhil Churamani, Atahan Çaldır, and Hatice Gunes. 2025. Participant Perceptions of a Robotic Coach Conducting Positive Psychology Exercises: A Qualitative Analysis. *J. Hum.-Robot Interact.* 14, 2, Article 36 (March 2025), 27 pages. doi:10.1145/3711937

[3] Dan Bennett, Oussama Metatla, Anne Roudaut, and Elisa D Mekler. 2023. How does HCI understand human agency and autonomy?. In *Proceedings of the 2023 CHI Conference on Human Factors in Computing Systems*. 1–18.

[4] Ryan K. Boettger and Laura A. Palmer. 2010. Quantitative Content Analysis: Its Use in Technical Communication. *IEEE Transactions on Professional Communication* 53, 4 (2010), 346–357. doi:10.1109/TPC.2010.2077450

[5] Umut Çakan, M Onur Keskin, and Reyhan Aydoğan. 2023. Effects of agent's embodiment in human-agent negotiations. In *Proceedings of the 23rd ACM International Conference on Intelligent Virtual Agents*. 1–8.

[6] Peter J Carnevale, Dean G Pruitt, et al. 1992. Negotiation and mediation. *Annual review of psychology* 43, 1 (1992), 531–582.

[7] Hyojin Chin, Hyeonho Song, Gumhee Baek, Mingi Shin, Chani Jung, Meeyoung Cha, Junghoi Choi, and Chiyoung Cha. 2023. The potential of chatbots for emotional support and promoting mental well-being in different cultures: mixed methods study. *Journal of Medical Internet Research* 25 (2023), e51712.

[8] Robert B Cialdini and Noah J Goldstein. 2004. Social influence: Compliance and conformity. *Annu. Rev. Psychol.* 55, 1 (2004), 591–621.

[9] George T Doran. 1981. There's a SMART way to write managements's goals and objectives. *Management review* 70, 11 (1981).

[10] Elizabeth Victoria Eikey, Clara Marques Caldeira, Mayara Costa Figueiredo, Yunan Chen, Jessica L Borelli, Melissa Mazmanian, and Kai Zheng. 2021. Beyond self-reflection: introducing the concept of rumination in personal informatics. *Personal and Ubiquitous Computing* 25, 3 (2021), 601–616.

[11] Friederike Eyssel, Ricarda Wullenkord, and Verena Nitsch. 2017. The role of self-disclosure in human-robot interaction. In *2017 26th IEEE international symposium on robot and human interactive communication (RO-MAN)*. IEEE, 922–927.

[12] Kathleen Kara Fitzpatrick, Alison Darcy, and Molly Vierhile. 2017. Delivering cognitive behavior therapy to young adults with symptoms of depression and anxiety using a fully automated conversational agent (Woebot): a randomized controlled trial. *JMIR mental health* 4, 2 (2017), e7785.

[13] Felix Glawe, Tim Schmeckel, Philipp Brauner, and Martina Ziefle. 2025. Human Autonomy and Sense of Agency in Human-Robot Interaction: A Systematic Literature Review. arXiv:2509.22271 [cs.HC] https://arxiv.org/abs/2509.22271

[14] Yijie Guo, Ruhan Wang, Zhenhan Huang, Tongtong Jin, Xiwen Yao, Yuan-Ling Feng, Weiwei Zhang, Yuan Yao, and Haipeng Mi. 2025. Exploring the Design of LLM-based Agent in Enhancing Self-disclosure Among the Older Adults. In *Proceedings of the 2025 CHI Conference on Human Factors in Computing Systems*. 1–17.

[15] Kerstin S Haring, Kelly M Satterfield, Chad C Tossell, Ewart J De Visser, Joseph R Lyons, Vincent F Mancuso, Victor S Finomore, and Gregory J Funke. 2021. Robot authority in human-robot teaming: Effects of human-likeness and physical embodiment on compliance. *Frontiers in Psychology* 12 (2021), 625713.

[16] Steven C Hayes, Jason B Luoma, Frank W Bond, Akihiko Masuda, and Jason Lillis. 2006. Acceptance and commitment therapy: Model, processes and outcomes. *Behaviour research and therapy* 44, 1 (2006), 1–25.

[17] Olivia Herzog, Niklas Forchhammer, Penny Kong, Philipp Maruhn, Henriette Cornet, and Fritz Frenkler. 2022. The Influence of Robot Designs on Human Compliance and Emotion: A Virtual Reality Study in the Context of Future Public Transport. *J. Hum.-Robot Interact.* 11, 2, Article 21 (March 2022), 17 pages. doi:10.1145/3507472

[18] Suzanne Janssen and Bob Schadenberg. 2024. A Psychological Need-Fulfillment Perspective for Designing Social Robots that Support Well-Being. *International Journal of Social Robotics* 16 (02 2024). doi:10.1007/s12369-024-01102-8

[19] Matouš Jelínek and Kerstin Fischer. 2021. The Role of a Social Robot in Behavior Change Coaching. In *Companion of the 2021 ACM/IEEE International Conference on Human-Robot Interaction* (Boulder, CO, USA) *(HRI '21 Companion)*. Association for Computing Machinery, New York, NY, USA, 434–438. doi:10.1145/3434074.3447208

[20] Sooyeon Jeong, Laura Aymerich-Franch, Sharifa Alghowinem, Rosalind W. Picard, Cynthia L. Breazeal, and Hae Won Park. 2023. A Robotic Companion for Psychological Well-being: A Long-term Investigation of Companionship and Therapeutic Alliance. In *Proceedings of the 2023 ACM/IEEE International Conference on Human-Robot Interaction* (Stockholm, Sweden) *(HRI '23)*. Association for Computing Machinery, New York, NY, USA, 485–494. doi:10.1145/3568162.3578625

[21] M Onur Keskin, Selen Akay, Ayse Dogan, Berkecan Koçyigit, Junko Kanero, and Reyhan Aydogan. 2024. You look nice, but i am here to negotiate: The influence of robot appearance on negotiation dynamics. In *Companion of the 2024 ACM/IEEE International Conference on Human-Robot Interaction*. 598–602.

[22] Taewan Kim, Seolyeong Bae, Hyun Ah Kim, Su-woo Lee, Hwajung Hong, Chanmo Yang, and Young-Ho Kim. 2024. MindfulDiary: Harnessing large language model to support psychiatric patients' journaling. In *Proceedings of the 2024 CHI Conference on Human Factors in Computing Systems*. 1–20.

[23] Kurt Kroenke, Robert L Spitzer, and Janet BW Williams. 2001. The PHQ-9: validity of a brief depression severity measure. *Journal of general internal medicine* 16, 9 (2001), 606–613.

[24] Guy Laban and Emily S. Cross. 2024. Sharing our Emotions with Robots: Why do we do it and how does it make us feel? *IEEE Transactions on Affective Computing* (2024), 1–18. doi:10.1109/TAFFC.2024.3470984

[25] Guy Laban, Val Morrison, Arvid Kappas, and Emily S. Cross. 2025. Coping with emotional distress via self-disclosure to robots: An intervention with caregivers. *International Journal of Social Robotics* (2025), 1–34.

[26] Guy Laban, Micol Spitale, Minja Axelsson, Nida Itrat Abbasi, and Hatice Gunes. 2025. Critical Insights about Robots for Mental Wellbeing. arXiv:2506.13739 [cs.RO] https://arxiv.org/abs/2506.13739

[27] Himanshi Lalwani and Hanan Salam. 2025. Supporting productivity skill development in college students through social robot coaching: A proof-of-concept. In *2025 34th IEEE International Conference on Robot and Human Interactive Communication (RO-MAN)*. IEEE, 1429–1436.

[28] Yi-Chieh Lee, Naomi Yamashita, Yun Huang, and Wai Fu. 2020. "I Hear You, I Feel You": Encouraging Deep Self-disclosure through a Chatbot. In *Proceedings of the 2020 CHI Conference on Human Factors in Computing Systems* (Honolulu, HI, USA) *(CHI '20)*. Association for Computing Machinery, New York, NY, USA, 1–12. doi:10.1145/3313831.3376175

[29] Han Li, Renwen Zhang, Yi-Chieh Lee, Robert E Kraut, and David C Mohr. 2023. Systematic review and meta-analysis of AI-based conversational agents for promoting mental health and well-being. *NPJ Digital Medicine* 6, 1 (2023), 236.

[30] Albert Łukasik and Arkadiusz Gut. 2025. From robots to chatbots: unveiling the dynamics of human-AI interaction. *Frontiers in Psychology* 16 (2025), 1569277.

[31] Bethany Nichol, Jemma McCready, Goran Erfani, Dania Comparcini, Valentina Simonetti, Giancarlo Cicolini, Kristina Mikkonen, Miyae Yamakawa, and Marco Tomietto. 2024. Exploring the impact of socially assistive robots on health and wellbeing across the lifespan: An umbrella review and meta-analysis. *International journal of nursing studies* 153 (2024), 104730.

[32] Susan Nolen-Hoeksema, Blair E Wisco, and Sonja Lyubomirsky. 2008. Rethinking rumination. *Perspectives on psychological science* 3, 5 (2008), 400–424.

[33] Dennis Papini, F Farmer, S Clark, J Micka, and J Barnett. 1990. Early adolescent age and gender differences in patterns of emotional disclosure to parents and friends. *Adolescence* 25 (02 1990), 959–76.

[34] Maria Pinto-Bernal, Matthijs Biondina, and Tony Belpaeme. 2025. Designing Social Robots with LLMs for Engaging Human Interaction. *Applied Sciences* 15, 11 (2025), 6377.

[35] Ilaria Riboldi, Angela Calabrese, Susanna Piacenti, Chiara Alessandra Capogrosso, Susanna Lucini Paioni, Francesco Bartoli, Giuseppe Carrà, Jo Armes, Cath Taylor, and Cristina Crocamo. 2024. Understanding university students' Perspectives towards digital tools for mental health support: A cross-country study. *Clinical Practice and Epidemiology in Mental Health: CP & EMH* 20 (2024), e17450179271467.

[36] Richard M Ryan and Edward L Deci. 2000. Self-determination theory and the facilitation of intrinsic motivation, social development, and well-being. *American psychologist* 55, 1 (2000), 68.

[37] Hanan Salam, Oya Celiktutan, Hatice Gunes, and Mohamed Chetouani. 2023. Automatic context-aware inference of engagement in hmi: A survey. *IEEE transactions on affective computing* 15, 2 (2023), 445–464.

[38] Zhonghao Shi, Ellen Landrum, Amy O'Connell, Mina Kian, Leticia Pinto-Alva, Kaleen Shrestha, Xiaoyuan Zhu, and Maja J Matarić. 2024. How can large language models enable better socially assistive human-robot interaction: a brief survey. In *Proceedings of the AAAI Symposium Series*, Vol. 3. 401–404.

[39] Dominic Ethan Sia, Marco Jalen Yu, Justine Leo Daliva, Jaycee Montenegro, and Ethel Ong. 2021. Investigating the Acceptability and Perceived Effectiveness of a Chatbot in Helping Students Assess their Well-being. In *Proceedings of the Asian CHI Symposium 2021* (Yokohama, Japan) *(Asian CHI '21)*. Association for Computing Machinery, New York, NY, USA, 34–40. doi:10.1145/3429360.3468177

[40] Xiaoxiao Sun, Adnan Jahangir, Abdulelah Ahmed Alghamdi, and Fang Liu. 2025. Use of ai-based mental health tools and psychological well-being among Chinese university students: a parallel mediation model of emotional self-efficacy and perceived autonomy. *Scientific Reports* 15, 1 (2025), 40305.

[41] Sandra Ulrich, Natascha Lienhard, Hansjörg Künzli, and Tobias Kowatsch. 2024. A chatbot-delivered stress management coaching for students (MISHA app): pilot randomized controlled trial. *JMIR mHealth and uHealth* 12 (2024), e54945.

[42] Edward Watkins. 2008. Constructive and Unconstructive Repetitive Thought. *Psychological Bulletin* 134 (03 2008), 163–206. doi:10.1037/0033-2909.134.2.163

[43] Ceng Zhang, Junxin Chen, Jiatong Li, Yanhong Peng, and Zebing Mao. 2023. Biomimetic Intelligence and Robotics. *Biomimetic Intelligence and Robotics* (2023), 100131.